\documentclass[conference]{IEEEtran}
\IEEEoverridecommandlockouts
\usepackage{cite}
\usepackage{amsmath,amssymb,amsfonts}
\usepackage{algorithmic}
\usepackage{graphicx}
\usepackage{textcomp}
\usepackage{xcolor}
\usepackage{booktabs}    
\usepackage{multirow}    
\usepackage{amsmath}     
\usepackage{array}       
\usepackage{caption}  
\usepackage{hyperref}
\def\BibTeX{{\rm B\kern-.05em{\sc i\kern-.025em b}\kern-.08em
    T\kern-.1667em\lower.7ex\hbox{E}\kern-.125emX}}
\begin{document}

\title{Benchmarking Cross-Scale Perception Ability of Large Multimodal Models in Material Science}

\author{
    \IEEEauthorblockN{Yuting Zheng\textsuperscript{1,2}, Zijian Chen\textsuperscript{1,2}\textsuperscript{*}, Jia Qi\textsuperscript{1}}
    
    \IEEEauthorblockA{
        \textsuperscript{1}Shanghai Artificial Intelligence Laboratory \\
        \textsuperscript{2}Shanghai Jiao Tong University \\
        Email: \{zhengyt058, zijian.chen\}@sjtu.edu.cn, jiaqi@pjlab.org.cn
    }
    \thanks{* Corresponding author.}
}

\maketitle


\begin{abstract}
Unraveling the hierarchical structure-property relationships is the central challenge of materials science, necessitating the interpretation of data across vast physical scales from micro to macro. Despite the rapid integration of Large Multimodal Models (LMMs) into scientific workflows, existing scientific benchmarks primarily focus on general chart interpretation or isolated common-sense reasoning, failing to capture reasoning ability across intricate physical dimensions. To address this, we introduce CSMBench, a dataset comprising 1,041 high-quality figures curated from premier journals up to September 2025. CSMBench categorizes data into four scientifically distinct regimes: atomic, micro, meso, and macro scales, strictly aligning with the focus and definitions in materials study. Through open-ended figure description and multiple-choice caption matching tasks, we evaluate state-of-the-art open-source and closed-source models. Our analysis identifies that performance varies significantly across physical scales due to the distinct visual characteristics, highlighting the limitations of current generalist models and identifying critical directions for achieving hierarchical and accurate understanding in materials research. The CSMBench is publicly released at: \url{https://huggingface.co/datasets/lututu/CSMBench} \end{abstract}




\section{Introduction}

\label{sec:intro}

Materials science serves as the primary engine for modern technological advancement, underpinning innovations in high-efficiency energy storage, aerospace structural integrity, and next-generation semiconductors~\cite{Accelerating}. This field is fundamentally defined by the hierarchical nature of matter, where the macroscopic performance of a material is the cumulative result of structures across multiple physical dimensions. From the arrangement of individual atoms to the distribution of grains at the micrometer scale and the final bulk geometry, these features evolve through complex and synergistic interactions. Consequently, the ability to perform cross-scale reasoning, which links microscopic observations to macroscopic outcomes, is the definitive hallmark of domain expertise in materials research.

As LMMs continue to progress from general-purpose understanding to domain-specific knowledge~\cite{zhang2025aibench, chen2025just}, scientific discovery has emerged as a critical frontier for evaluating and extending their abilities~\cite{sciencellm1}. Recent advances in LMMs have achieved remarkable performance across diverse benchmarks, yet their effectiveness in the specialized materials domain remains largely unproven. While existing scientific benchmarks such as ScienceQA~\cite{scienceqa} evaluate general science topics or college-level reasoning, they often rely on secondary sources such as textbooks and academic materials~\cite{mathvista}. Consequently, these generalist benchmarks fail to capture the multiscale nature of real-world materials research. They do not evaluate whether a model can traverse the physical dimensions from angstrom-level lattices to centimeter-scale bulk materials, leaving a fundamental challenge in granularly measuring scientific cognitive capabilities across hierarchical physical regimes.

To bridge this critical gap, we introduce CSMBench, a {\bf C}ross-{\bf S}cale {\bf M}aterial science {\bf Bench}mark for LMMs, which is designed to evaluate the hierarchical understanding of LMMs through four physical regimes. The atomic scale characterizes the capacity to discern lattice arrangements and individual atomic defects. The microscale demonstrates the ability to interpret nanometer-to-micrometer features such as precipitates and dislocations. The mesoscale manifests the ability to analyze micrometer-to-millimeter structures, including grain boundaries and textures. The macro scale represents the understanding of centimeter-scale bulk materials and final component geometries.

We benchmark state-of-the-art proprietary and  open-source LMMs using CSMBench, as illustrated in Table~\ref{tab:main-results-openqa}. These models exhibit suboptimal results on our framework, which indicates that CSMBench serves as a challenging frontier for scientific LMMs development, exposing the gap between visual perception and deep physical reasoning. Our core contributions are summarized as follows:

\begin{enumerate}

\item We curate a contamination-free dataset of 1,041 figures sourced from the latest 2025 peer-reviewed literature, ensuring that models are evaluated on novel data beyond their training cutoff.

\item We categorize the dataset into four physical scales, including atomic, micro, meso, and macro, and design two tasks, including open-ended figure description and multiple-choice caption matching, to measure both recognition precision and reasoning depth.

\item We comprehensively evaluate ten state-of-the-art LMMs, revealing that performance varies significantly across physical scales due to the distinct visual characteristics.

\end{enumerate}

\begin{figure*}[t]
\centerline{\includegraphics[scale=0.56]{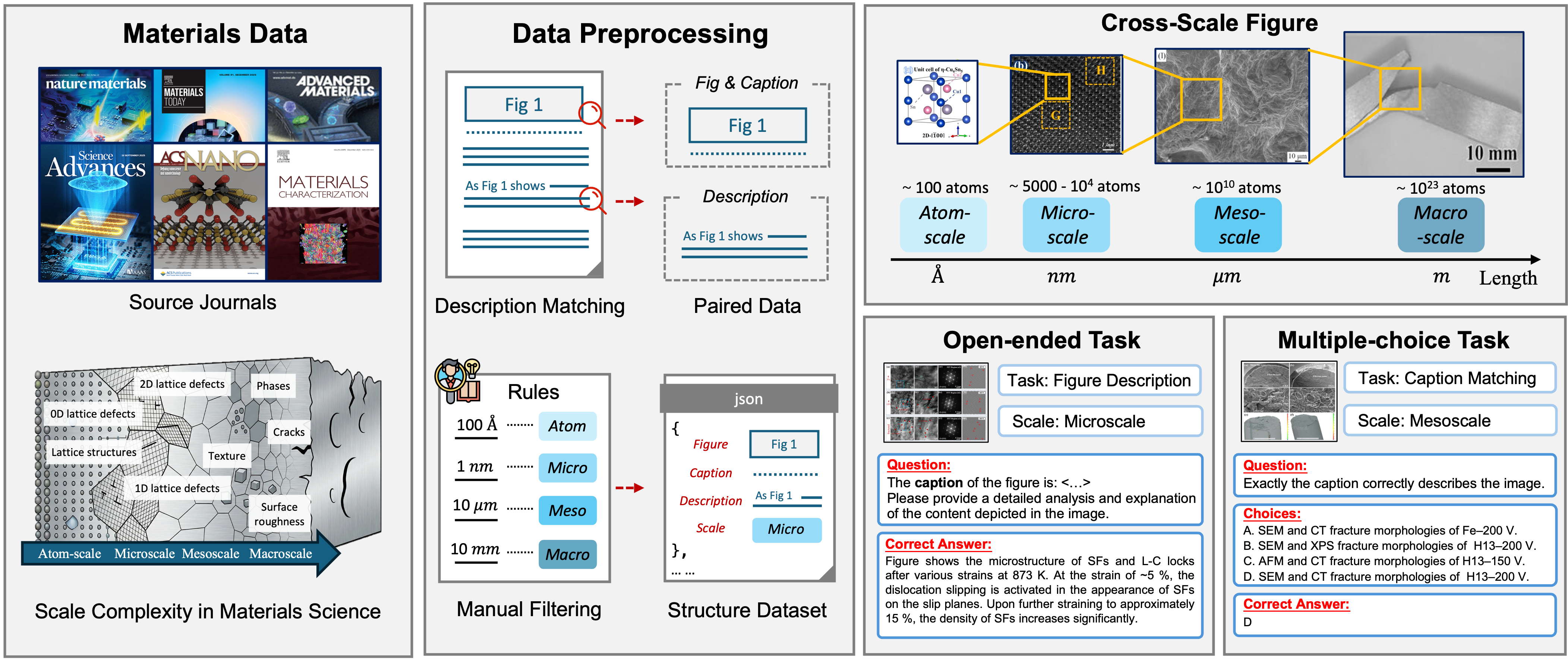}}
\caption{ An overview of CSMBench, which consists of three primary phases: source data collection, data processing, and task design. The pipeline aggregates figures from authoritative materials journals across four physical scales, ranging from the atomic level to the macroscopic level, which strictly adhere to fundamental materials science definitions. The bottom-left schematic\protect\footnotemark ~highlights the intrinsic characteristics across scales, serving as the theoretical basis for our scale categorization. Through a hybrid workflow combining automated description matching and expert manual filtering, raw data is refined into a high-quality structured dataset. Finally, the framework establishes two evaluation tasks to assess cross-scale perception: open-ended figure description and multiple-choice caption matching.}
\label{hybird}
\end{figure*}
\section{Related Work}
\label{sec:related}

{\it Science LMMs}:
The integration of domain knowledge into LMMs has become a cornerstone of modern scientific discovery, driving a shift from simple text processing to comprehensive visual understanding. This evolution is evident in fields like chemistry and biomedicine, where models such as and Progen~\cite{progen} utilize specialized knowledge to decipher complex sequences, and in geosciences, where EarthGPT~\cite{earthgpt} integrate multimodal satellite data for environmental analysis. Following this trajectory, materials science has seen a surge in LMMs applications, with innovations like MatterChat~\cite{matterchat} which unifies crystal structures with text, and HoneyComb~\cite{honeycomb} which establishes agentic workflows for research. However, unlike sequence-based domains, materials science is fundamentally governed by visual features across disparate physical scales. Although recent works have incorporated multimodal capabilities, the primary challenge lies in enabling these models to interpret not only isolated images but also the hierarchical physical laws inherent within them.

\vspace{-4pt} 
{\it Science Benchmarks}:
Recent efforts to evaluate LMMs in the scientific sphere have led to the creation of benchmarks like ScienceQA~\cite{scienceqa}, which consists of $\sim$21k multimodal multiple-choice questions on diverse science topics. Specialized benchmarks such as MathVista~\cite{mathvista}  further challenge models with mathematical visualizations. Within the scientific figure understanding domain, MMSCI~\cite{mmsci} utilizes figures from \textit{Nature Communications} across 72 fields, but its evaluation remains centered on general chart interpretation rather than deep domain expertise. Similarly, MaCBench~\cite{macbench} introduced a multimodal evaluation for materials science, yet it is largely confined to common-sense interpretations. Importantly, these generalist benchmarks fail to capture the multi-scale nature of materials and evaluate the ability to traverse the physical dimensions from atom-scale to centimeter-scale bulk materials, leaving the community without a rigorous benchmark for scale-aware reasoning.

\begin{table*}[t] 
\centering
\caption{Performance comparison of open-ended question answering across four scales: atom scale, microscale, mesoscale, and macroscale. Reported metrics include BERTScore (F1),  Semantic Textual Similarity (STS), and LLM-as-a-Judge score (LLM). All metrics follow a higher-is-better trend ($\uparrow$). Bold entries denote the best results.}
\label{tab:main-results-openqa}
\begin{tabular}{l >{\centering\arraybackslash}p{0.62cm} >{\centering\arraybackslash}p{0.62cm} >{\centering\arraybackslash}p{0.62cm} 
              >{\centering\arraybackslash}p{0.62cm} >{\centering\arraybackslash}p{0.62cm} >{\centering\arraybackslash}p{0.62cm}
              >{\centering\arraybackslash}p{0.62cm} >{\centering\arraybackslash}p{0.62cm} >{\centering\arraybackslash}p{0.62cm}
              >{\centering\arraybackslash}p{0.62cm} >{\centering\arraybackslash}p{0.62cm} >{\centering\arraybackslash}p{0.62cm}
              >{\centering\arraybackslash}p{0.62cm} >{\centering\arraybackslash}p{0.62cm} >{\centering\arraybackslash}p{0.62cm}}
\toprule
\multirow{2}{*}{Model} 
& \multicolumn{3}{c}{AtomScale} 
& \multicolumn{3}{c}{Microscale} 
& \multicolumn{3}{c}{Mesoscale} 
& \multicolumn{3}{c}{Macroscale} 
& \multicolumn{3}{c}{Overall}\\
\cmidrule(lr){2-4} \cmidrule(lr){5-7} \cmidrule(lr){8-10} \cmidrule(lr){11-13} \cmidrule(lr){14-16} 

& Bert & STS & LLM 
 & Bert & STS & LLM
& Bert & STS & LLM 
 & Bert & STS & LLM 
 & Bert & STS & LLM\\
\midrule
GPT-5.1    & 0.6101 & 0.7006  & \textbf{7.8679} & 0.6129 & \textbf{0.6917} & \textbf{8.2672} & 0.6613 & \textbf{0.7011 }& \textbf{8.0971} & 0.5975 & 0.6613 & \textbf{7.5217} & 0.6205 & 0.6887 & \textbf{7.9385} \\
Gemini-2.5-pro   & \textbf{0.6268} & \textbf{0.7280} & 7.0566  & \textbf{0.6228} & 0.6777 & 7.5854 & \textbf{0.6232} & 0.6949 & 7.4429 & \textbf{0.6101} & \textbf{0.6599} & 6.7935 & \textbf{0.6207} & \textbf{0.6901} & 7.2196 \\
Doubao-1.6-vison  & 0.5837 & 0.6991 & 7.2075 & 0.5851 & 0.6794 & 7.6842 & 0.5829 & 0.6874 & 7.5593 & 0.5755 & 0.6530  & 7.0000 & 0.5818 & 0.6797 & 7.3628 \\

\midrule
Qwen2.5-VL-7B    & \textbf{0.6273} & 0.7169 & 5.2264 &  0.6322 & 0.7026 & 6.3806 & 0.6234 & 0.6975 &  5.8490 & 0.6040 & 0.6533 & 5.5217 & 0.6217 & 0.6926 & 5.7444 \\
Qwen2.5-VL-32B     &  0.5827 & 0.6660 & 4.5849 & 0.5890 & 0.6595 & 5.1174 & 0.5870 &  0.6664 & 5.1032 & 0.5720 & 0.6274 & 4.7065 & 0.5827 & 0.6548 & 4.8780 \\
Qwen2.5-VL-72B   &  0.6245 & \textbf{0.7199} & \textbf{6.5849} &  \textbf{0.6323} & \textbf{0.7055} & \textbf{6.9838} & \textbf{0.6291} & \textbf{0.7074} & 6.5485 & \textbf{0.6102} & \textbf{0.6613} & \textbf{5.8587} & \textbf{0.6240} & \textbf{0.6985} & \textbf{6.4940} \\

Qwen3-VL-8B & 0.6215& 0.7183 & 6.3774 & 0.6238 & 0.7028 &6.9069 & 0.6208 & 0.7040 &  \textbf{6.6256} & 0.6037 & 0.6559 & 5.8043 & 0.6175 & 0.6953 & 
6.4286 \\

InternVL3-8B    & 0.6108 & 0.6874 & 4.7358 & 0.6122  & 0.6867 & 5.3482 & 0.6082 & 0.6731 & 5.0277 & 0.5967 & 0.6413 & 4.8043 & 0.6070 & 0.6721 & 4.9790 \\
InternVL3-38B    & 0.6125 & 0.7045  & 5.7170 & 0.6100  & 0.6846 & 6.2186 & 0.6082 & 0.6839 & 5.9353 & 0.6009 & 0.6460 & 5.6848 & 0.6079 & 0.6798 & 5.8889 \\
InternVL3-78B    &  0.6142  & 0.7094 & 5.6792 & 0.6134  & 0.6824 & 6.3968 & 0.6084 & 0.6825 & 6.0139 & 0.6009 & 0.6446 & 5.7935 & 0.6092 & 0.6797 & 5.9709\\
		
\bottomrule
\end{tabular}
\end{table*}

\footnotetext{The schematic visualizes lattice defect types in solids that are affected by compositional complexity. The figure is adapted from the materials design work~\cite{Accelerating} which was published in nature computational science 2023.}
\section{CSMBench}
\label{dataset}
\subsection{Data Collection}
To ensure the benchmark is authoritative and representative of cross-scale characteristics, a rigorous data collection and processing pipeline is developed as follows.

\subsubsection{Source Data Collection}
The CSMBench dataset is derived from eight prominent materials science journals recognized for their authority in the field: Nature, Nature Materials, Nature Communications, Science Advances, Advanced Materials, Materials Today, and Materials Characterization. The inclusion of Materials Characterization is particularly notable for its emphasis on the structural behavior of materials and its extensive use of diverse characterization techniques. Articles included in the dataset span up to September 2025, ensuring the dataset remains uncontaminated by outdated data, thereby providing a reliable foundation for evaluating contemporary LMMs. A total of 1,041 figures across four distinct physical scales are curated from 432 peer-reviewed materials science articles. articles.

Following materials informatics principles, we categorize material research scales into four levels based on physical dimensions. The collected data cover the following four scales:

\begin{itemize}

\item \textbf{Atomic scale:} At this $\text{\AA}$ scale ($10^{-10}$m), materials research focuses on atomic-level phenomena like electronic structures, defects, and interfaces, which are foundational for understanding atomic bonding and interactions that define material properties.

\item \textbf{Micro-scale:} At the nanometer scale ($10^{-9}$m), studies investigate phenomena like diffusion, interface evolution, and material degradation, influencing properties like strength and conductivity at the grain or particle level.

\item \textbf{Meso-scale:} At the micrometer scale ($10^{-6}$m), research examines microstructural features such as phase boundaries, pores, and cracks. This scale bridges atomistic models with macroscopic mechanics, affecting electrochemical and mechanical properties.

\item \textbf{Macro-scale:} At the macroscopic scale (centimeters to meters), materials are considered as continuous media, with a focus on bulk behaviors such as deformation, fluid flow, and heat transfer in engineering applications.

\end{itemize}

\subsubsection{Data Preprocessing}
To obtain detailed explanations for each figure, which will serve as the ground truth to assess perception ability, we first convert raw PDFs into machine-readable formats using MinerU, which segments figures, converts formulas to LaTeX, and extracts text into Markdown. Then we develop a regular expression matching function to identify figure references (e.g., Fig. n) within the body text. Since this initial matching often identifies multiple candidate paragraphs for one figure, we implement a cleaning pipeline to exclude paragraphs that reference multiple figures or contain external table data. Following this refinement, we merge fragmented descriptions belonging to the same figure into a single unified paragraph. We then retained descriptions within a 100–300 token range to eliminate long-tail data. 

\subsubsection{Human Filtering}
To further ensure that the image data meets our cross-scale requirements, we conduct an expert manual review to filter the automatically extracted image data that only contain generic flowcharts or pure statistical charts and retain those that provide direct insights into material morphology, phase, or composition. Following this filtering, we perform scale annotation by manually verifying scale bars and contextual metadata for every image. This process results in 1,041 accurately categorized figures for evaluating cross-scale reasoning: 53 at the atomic scale, 247 at the micro-scale, 649 at the meso-scale, and 92 at the macro-scale. The uneven distribution of data is due to the inherent distribution of the samples themselves, which may be influenced by the difficulty of characterization and the focus of mechanism research.

\subsection{Task Design}

We develop two tasks in CSMBench with distinct settings to comprehensively evaluate the ability of LMMs to interpret materials science figures.

\subsubsection{Open-ended Task}
In this task, we evaluate the model ability of figure description, where the model receives a figure and its caption as input and is tasked to generate a detailed explanation within 100 to 300 words. The model is required to focus on the visible features of the image and draw scientific conclusions based on the presented evidence. The generated content is evaluated by comparing its similarity to high-fidelity ground-truth paragraphs extracted from the original articles.

This descriptive task is specifically designed to assess cross-scale perception ability. Unlike generic chart-to-text tasks, interpreting materials science figures requires a fine-grained understanding of both local and global features. For instance, a model must correctly identify atomic arrangements at the atomic scale, interface evolutions at the microscale, microstructural morphologies at the mesoscale, and bulk deformation patterns at the macroscale. By requiring models to analyze these visual cues, this task effectively examines whether a model possesses a deep, physically grounded understanding of materials science.

\subsubsection{Multiple-choice Task}
The second task format is multiple-choice and is designed to evaluate the caption matching task, where models are required to select the correct caption for a given figure from a set of four options, with fine-grained distractors included. To rigorously test the precision ability, we generate three types of challenging negative distractors through targeted perturbations, utilizing GPT-4.1 to introduce variations in the following ways:
    
\begin{itemize}

\item \textbf{Characterization Methods:} We prompt the model to replace the original technique with a similar but incorrect method (e.g., swapping TEM for SEM or XRD for XPS). This checks whether the model can recognize the specific visual signatures of different experimental instruments.

\item \textbf{Material Compositions:} We instruct the model to substitute specific material names or chemical formulas with plausible alternatives (e.g., $ZrB_2$ to $TiB_2$, or $Al_2O_3$ to $Fe_2O_3$). This evaluates the model's ability to connect textual chemical knowledge with the visual context.

\item \textbf{Numerical Values:} We have the model perturb critical parameters such as temperatures, concentrations, dimensions, and time (e.g., 1050°C to 950°C, or 2 h to 4 h). Accurate numerical recognition is crucial for precise scientific reasoning.

\end{itemize}

By forcing the model to distinguish between these subtle variations, we evaluate its ability to perform high-precision recognition of characterization methods, compositions, and quantitative data, all of which are fundamental to the correct interpretation of materials research.

\begin{table}[t] 
\centering
\caption{Performance comparison of different models on multiple-choice questions across the four scales. We report the accuracy scores. The best results are highlighted in bold.}
\label{tab:main-results-mcqa}
\begin{tabular}{l >{\centering\arraybackslash}p{0.85cm} >{\centering\arraybackslash}p{0.85cm} >{\centering\arraybackslash}p{0.85cm} 
              >{\centering\arraybackslash}p{0.85cm} >{\centering\arraybackslash}p{0.85cm} >{\centering\arraybackslash}p{0.85cm}}
\toprule
 Model   & Atom.  & Micro.  & Meso.  & Macro. & Overall \\
\midrule
GPT-5.1  & 0.8491 & 0.9271 & 0.8829 & 0.9565 & 0.9039 \\
Gemini-2.5-pro   & 0.9434 & 0.9433 & 0.9137 &  0.9348  & 0.9338\\
Doubao-1.6-vison  & \textbf{0.9434} & \textbf{0.9636} & \textbf{0.9492} & \textbf{0.9674} & \textbf{0.9559}\\
\midrule
Qwen2.5-VL-7B   & 0.6604 & 0.7895 & 0.7458 & 0.7065 & 0.7256 \\
Qwen2.5-VL-32B   & 0.7170 & 0.7814 & 0.7935 & 0.7935 & 0.7714 \\
Qwen2.5-VL-72B   & 0.7170 & 0.8300 & 0.8136 & \textbf{0.8587}  & 0.8048 \\
Qwen3-VL-8B  & \textbf{0.8626} & \textbf{0.9057} & \textbf{0.8988} & 0.8536 & \textbf{0.8802} \\
InternVL3-8B   & 0.6981 & 0.7166 & 0.7057 & 0.6304 & 0.6877 \\
InternVL3-38B  & 0.7736 & 0.8866 & 0.8582 & 0.8478 & 0.8416 \\
InternVL3-78B  & 0.7736 & 0.8421 & 0.8320 & 0.8152 & 0.8157 \\
\bottomrule
\end{tabular}
\end{table}

\section{Experiment}

\label{sec:exp}

\subsection{Experimental Setup}

\subsubsection{Benchmark Candidates}
To establish a comprehensive performance baseline on CSMBench, we evaluate a diverse array of Large Multimodal Models (LMMs). Our selection comprises a mix of cutting-edge proprietary and representative open-source models. Specifically, for proprietary models, we include state-of-the-art candidates such as GPT-5.1~\cite{openai2025gpt}, Grok-4~\cite{xai2025grok4fast}, and Gemini-2.5-Pro~\cite{comanici2025gemini}. For open-source LMMs, we focus on models recognized for their robust multimodal capabilities, specifically the Qwen-2.5-VL-\{7, 32, 72\}B~\cite{qwen25}, Qwen-3-VL-8B~\cite{qwen3technicalreport} and InternVL3-\{8, 38, 78\}B~\cite{zhu2025internvl3}.

\subsubsection{Metrics}
For the multiple-choice QA task, we present accuracy as the primary metric. For the open-ended QA task, we report three metrics. For \textbf{BERTScore}~\cite{bertscore}, we employ the F1 score to evaluate semantic similarity using contextual embeddings. \textbf{Semantic Textual Similarity (STS)} is used to provide a high-level conceptual assessment of the overall interpretation. We calculate STS by projecting both the model output and ground truth into a unified 384-dimensional vector space using a sentence-transformer model~\cite{sms-minilm}, and computing their cosine similarity. For the \textbf{LLM-as-a-Judge score}, we use GPT-4o-2024-11-20~\cite{openai2025gpt} as the judge to semantically verify the correctness of answers against model predictions, following the implementation and prompt developed in the previous work~\cite{zhou2025scientists}, where the score is scaled from 1 to 10. The use of LLM score together with STS helps exclude family bias where GPT-4o may favor models from the GPT family. The results are considered reliable when scores are consistent.

\begin{figure}[t]
\centerline{\includegraphics[scale=0.24]{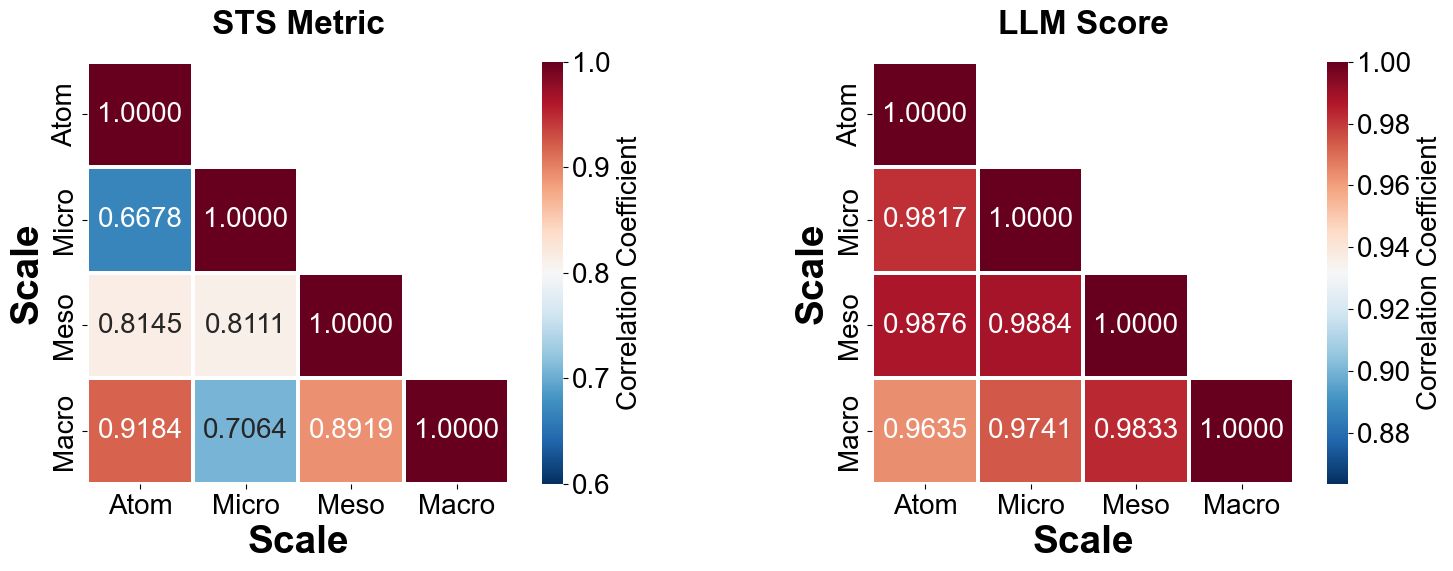}}
\caption{Pearson correlation coefficients between the atom, micro, meso, and macro scales using the STS metric and LLM score representatively.}
\label{fig: correlation}
\end{figure}

\subsection{Main Results}

\textbf{Observation 1. Proprietary models perform better in both scientific reasoning and discriminative accuracy.} Among all evaluated candidates, proprietary models demonstrate a systematic advantage. As shown in Table.~\ref{tab:main-results-openqa}, GPT-5.1 achieves the highest overall LLM-as-a-Judge score of 8.0749 in the open-ended description task, indicating a superior ability to generate physically grounded and scientifically accurate narratives. In the multiple-choice matching task as shown in Table.~\ref{tab:main-results-mcqa}, Doubao-1.6-vision emerges as the top performer with a near-perfect overall accuracy of 95.59\%, followed by Gemini-2.5-pro at 93.38\%.
A significant performance gap remains between proprietary and open-source models. In the open-ended task as shown in Table.~\ref{tab:main-results-openqa}, the strongest open-source representative, Qwen2.5-VL-72B, reaches an overall LLM score of 6.4940, which trails GPT-5.1 by approximately 1.82 points. This gap is similarly pronounced in the fine-grained matching task, where even the most capable open-source models often lag behind the leading proprietary models by a margin of nearly 8\% in overall accuracy. These results confirm that while open-source models are progressing, proprietary systems still maintain a stronger grasp of the complex, multi-modal reasoning required for materials science.

\textbf{Observation 2. LMMs exhibit scale-dependent performance fluctuations.} 
Analysis of results across the four physical scales reveals that models generally perform better at the microscale and mesoscale than at the atomic and macro scales. GPT-5.1 achieves a peak LLM-as-a-judge score of 8.2672 at the microscale, which significantly drops to 7.5217 at the macroscale. This trend is mirrored by open-source models such as Qwen3-VL-8B, where the score declines from 6.9069 at the microscale to 5.8043 at the macroscale. This performance disparity is primarily attributed to the nature of the visual data within each scale. The micro and meso scales are dominated by standardized images from experimental characterization, including SEM, CT, and XPS. These images possess relatively consistent visual patterns that models can interpret with greater ease. In contrast, datasets for the atomic and macro scales contain a higher proportion of diverse schematic diagrams and heterogeneous visual representations, which increase the complexity of the task of interpretation. Furthermore, the correlation analysis shown in Fig.~\ref{fig: correlation} confirms this scale dependency. Although the scores for the LLM-as-a-judge metric exhibit great cross-scale consistency, exceeding 0.97, which may be attributed to the stable internal logic of a model, the STS metric shows significantly higher variance and drops to 0.6678. This indicates that the precision of semantics and terminology is highly sensitive to the specific visual context and the scale-dependent language of materials science.

\begin{figure}[t]
\centering
\includegraphics[width=\columnwidth]{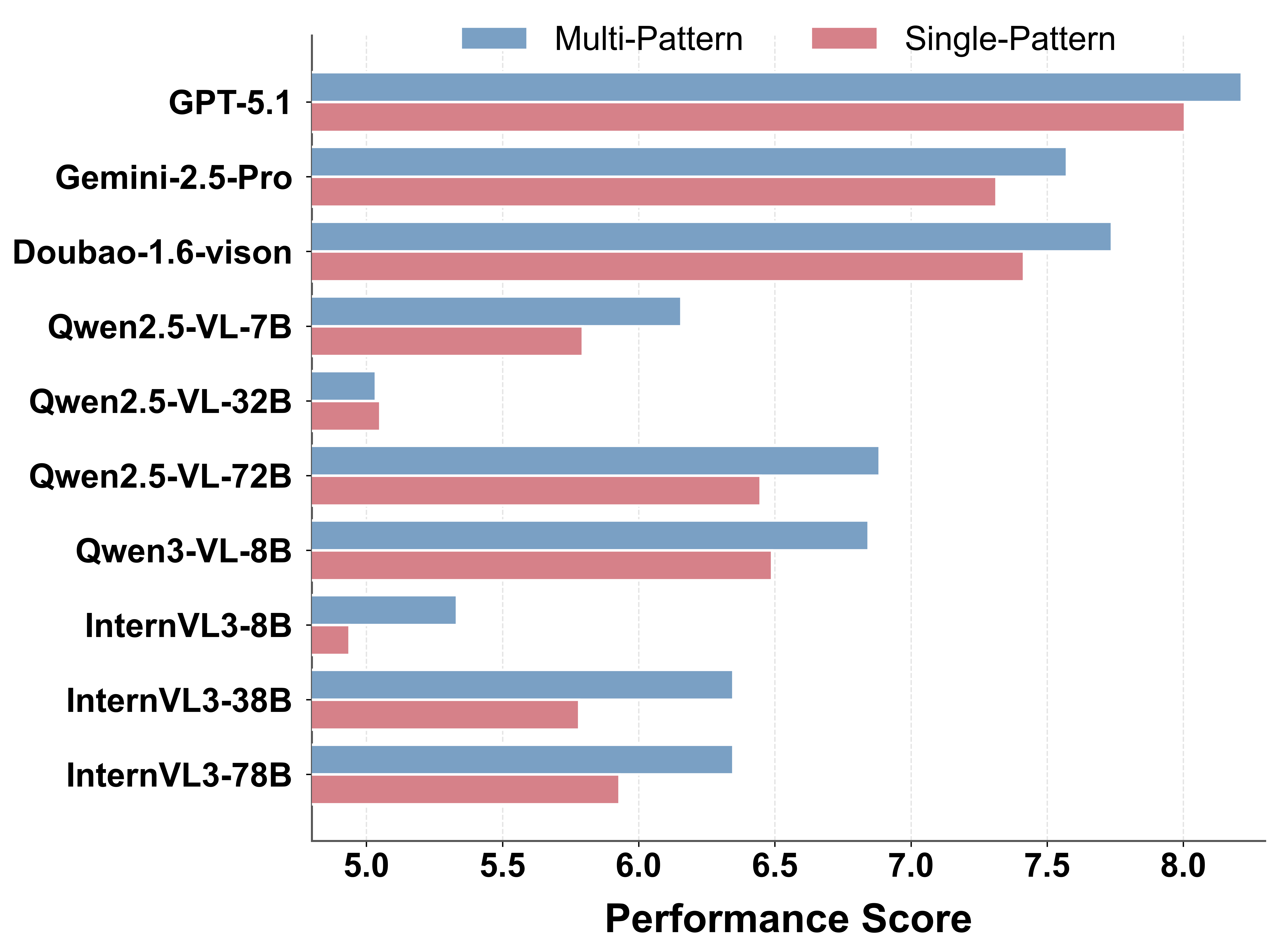}
\caption{Performance comparison on pure morphological images and charted hybrid multi-pattern images.}
\label{fig: hybrid}
\end{figure}

\textbf{Observation 3. Reasoning capabilities are prior to the scaling laws in materials science interpretation.} We can observe the scaling laws from Fig.~\ref{tab:main-results-mcqa} that within the Qwen2.5-VL family, matching accuracy improves from 72.56\% for the 7B model to 80.48\% for the 72B version. However, this scaling is inconsistent in complex tasks. As shown in Fig.~\ref{tab:main-results-openqa}, the 32B model records a score of 4.8780 in the open-ended figure description task and fails to surpass the 5.7444 achieved by the 7B model. Similarly, InternVL3-38B reaches 84.16\% accuracy and exceeds the 81.57\% of the larger 78B variant in the multiple-choice caption matching task. These findings indicate that simply increasing model parameters does not ensure stronger semantic integration in scientific contexts. Additionally, the performance of Qwen3-VL-8B further highlights the value of architectural evolution. This model achieves a score of 88.02\% and outperforms the much larger Qwen2.5-VL-72B, which scores 80.48\%.  By incorporating an internal thinking process, Qwen3-VL-8B can systematically analyze the hierarchical features of materials science images, which enables a more sophisticated alignment between visual evidence and physical principles.

\vspace{2pt}
\textbf{Observation 4. Discrepancy between open-ended description task and multiple-choice caption matching task. }
A direct comparison of the two tasks reveals that some models excel at identifying the correct choice without being able to articulate the underlying domain information in the figure. Most notably, Doubao-1.6-vision achieves the highest overall accuracy of 95.59\% in the matching task, demonstrating an exceptional ability to discriminate between fine-grained captions. However, in the open-ended figure description task, it is surpassed by GPT-5.1, which maintains a superior LLM-as-a-judge score of 7.9385. Similarly, among open-source models, InternVL3-38B exhibits strong discriminative performance with 84.16\% matching accuracy but is notably outperformed in scientific narrative by Qwen3-VL-8B, which maintains a superior LLM score of 6.4286. These results suggest that many models rely on visual heuristics to solve multiple-choice questions while lacking the cross-scale structural mapping and physical property derivation necessary for high-quality scientific description.

\begin{figure}[t]
\centering
\includegraphics[width= \columnwidth]{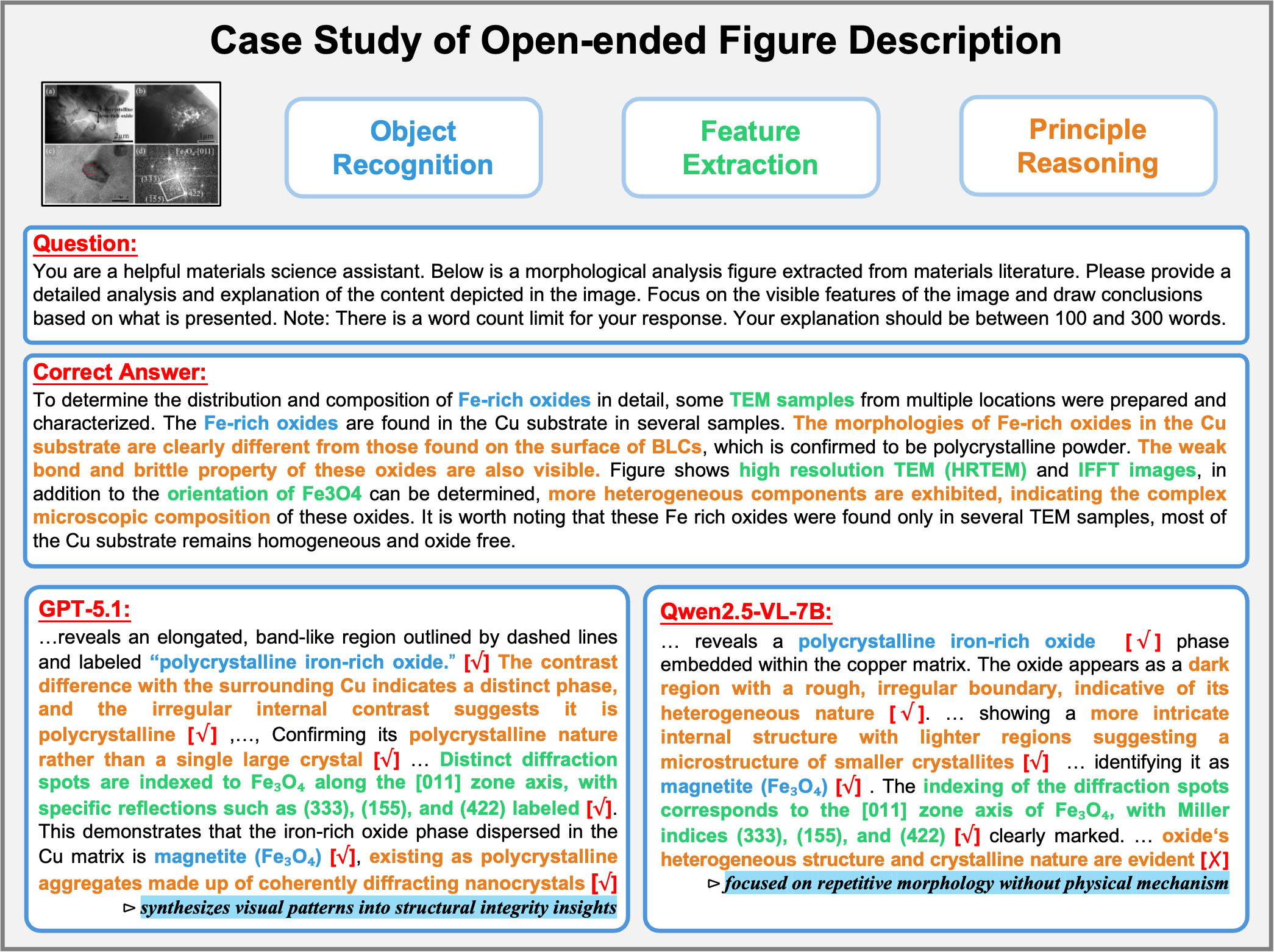}
\vspace{-3ex}
\caption{A comparative case study of GPT-5.1 and Qwen2.5-VL-7B across the dimensions of object recognition, feature extraction, and principle reasoning.}
\label{fig: erro}
\end{figure}


\textbf{Observation 5. Heterogeneous visual patterns strengthen interpretive performance.} Visual data in materials science primarily consist of morphological characterizations that capture physical structures and auxiliary statistical plots that present property-related information or statistical analysis. To evaluate the impact of visual diversity, we compared the performance of the models across two distinct categories of data: single morphological pattern and multi-patterns that integrate the two figure types. We calculated the scores for each category based on the LLM score metric, as shown in Fig.~\ref{fig: hybrid}. A majority of the models, including GPT-5.1, Gemini-2.5-Pro, and Qwen2.5-VL-72B, achieve higher scores on hybrid visual configurations. This improvement suggests that auxiliary statistical plots provide critical contextual anchors, which assist in the disambiguation of the complex structures of materials by providing complementary quantitative data. 

\subsection{Case Study} 
To investigate qualitative reasoning, we compare GPT-5.1 and Qwen-3-VL-8B across three levels of capability, specifically object recognition as exemplified by the identification of phases, feature extraction such as the parsing of technical signatures, and principle reasoning including the inference of physical mechanisms. The analysis is shown in Fig.~\ref{fig: erro}.
Both models demonstrate proficiency in object recognition and feature extraction by correctly identifying magnetite precipitates of the form $Fe_3O_4$ and extracting crystallographic signatures such as Miller indices. These results indicate a parity of performance in the domain of low-level visual perception. However, a significant gap remains in the area of principle reasoning. While the ground truth identifies the brittle nature and weak structural bonding of the oxides, the open-source model remains focused on repetitive descriptions of morphology. Conversely, the proprietary model excels by synthesizing visual data into a higher-level narrative regarding structural integrity. These results confirm that while the gap in perception between the proprietary model and the open-source model is closing, a gap in reasoning persists, as proprietary models show superiority at translating visual patterns into abstract physical principles.


\section{Conclusion}
\label{sec:conclu}
In this work, we propose CSMBench, which is designed to evaluate the hierarchical understanding of models across four physical scales in the study of materials. We curate 1041 high-quality figures from peer-reviewed literature up to September 2025 to provide a granular assessment of scientific perception. We reveal that performance varies significantly across physical scales due to the distinct visual characteristics. We further identify a critical gap where high recognition accuracy fails to translate into deep physical reasoning. We conclude that CSMBench serves as a challenging frontier to facilitate advancements in the development of physically grounded models for scientific discovery.

\section{Acknowledgment}
This work was supported by New Generation Artificial Intelligence-National Science and Technology Major Project (2025ZD0124104) in collaboration with Shanghai Artificial Intelligence Laboratory.

\bibliographystyle{IEEEbib}
\bibliography{icme2026references}

\end{document}